\documentclass[twocolumn,aps,pra,superscriptaddress,10pt]{revtex4-2}
\usepackage{graphicx}
\usepackage{amsmath, amsfonts, amssymb}
\usepackage{braket}
\usepackage{bm}
\usepackage{dcolumn}
\usepackage{subfigure}
\usepackage{graphicx}
\usepackage[colorlinks=true,linkcolor=blue,citecolor=blue,urlcolor=blue]{hyperref}
\usepackage{soul}
\usepackage{changes}
\usepackage{amsthm,amsmath,amssymb}
\usepackage{algorithm}
\usepackage{algorithmic}
\usepackage{mathrsfs}
\usepackage{multirow}

\makeatother

\begin{document}
%\begin{CJK*}{GB}{} 
%,twoside

\title{Wehrl Entropy and Entanglement Complexity   of Quantum  Spin Systems}%in an atomic heteronuclear spinor Bose gas}%of Ultracold Spin Mixture }

\author{Chen Xu}
%\email{Present address: Homer L. Dodge Department of Physics and Astronomy, The University of Oklahoma, 440 W. Brooks Street, Norman, Oklahoma 73019, USA}

\affiliation{Department of Physics, Renmin University of China, Beijing, 100872,
China}

\author{Yiqi Yu}
%\email{Present address: Homer L. Dodge Department of Physics and Astronomy, The University of Oklahoma, 440 W. Brooks Street, Norman, Oklahoma 73019, USA}

\affiliation{Department of Physics, Renmin University of China, Beijing, 100872,
China}
%\author{Dazhi Xu}
%\email{dzxu@bit.edu.cn}
%
%\affiliation{Center for Quantum Technology Research and School of Physics, Beijing
%Institute of Technology, Beijing 100081, China}

\author{Peng Zhang}
\email{pengzhang@ruc.edu.cn}

\affiliation{Department of Physics, Renmin University of China, Beijing, 100872,
China}

\affiliation{Key Laboratory of Quantum State Construction and Manipulation (Ministry of Education), Renmin University of China, Beijing, 100872, China}

\affiliation{Beijing Key Laboratory of Opto-electronic Functional Materials \&
Micro-nano Devices (Renmin University of China)}

\date{today}

\begin{abstract}

The Wehrl entropy  of  a  quantum state is the Shannon entropy of its coherent-state distribution function, and remains non-zero even for pure states. We investigate  
the relationship between this entropy
and the many-particle quantum entanglement,
for
 $N$ spin-1/2 particles.
Explicitly, we
numerically calculate the 
Wehrl entropy of various $N$-particle  ($2\leq N\leq 20$) entangled pure states,
 with respect to  the SU(2)$^{\otimes N}$ coherent states. Our results show that
for the large-$N$   ($N\gtrsim 10$)  systems the Wehrl entropy  of the highly chaotic entangled states ({\it e.g.}, $2^{-N/2}\sum_{s_1,s_2,...,s_N=\uparrow,\downarrow}|s_1,s_2,...,s_N\rangle e^{-i\phi_{s_1,s_2,...,s_N}}$, with $\phi_{s_1,s_2,...,s_N}$ being random angles) are substantially larger than that of the very regular 
entangled states ({\it e.g.}, the Greenberger–Horne–Zeilinger state). 
Therefore,  the Wehrl entropy can reflect
the complexity of the quantum entanglement
of many-body pure states,
as proposed by A. Sugita (Jour.
Phys. A {\bf 36}, 9081 (2003)). 
In particular,
 the  Wehrl entropy per particle (WEPP) can be used as a 
 quantitative description of this entanglement complexity.
Unlike  other quantities   used to evaluate this complexity ({\it e.g.}, the degree of entanglement between a subsystem and the other particles),  the WEPP does not necessitate the division of the total system into two subsystems. We further demonstrate that many-body pure entangled states can be classified into three types, based on the behavior of the WEPP in the limit $N \rightarrow \infty$: states approaching that of a maximally mixed state,  those approaching completely separable pure states, and a third category lying between these two extremes. Each type exhibits fundamentally different entanglement complexity.

\end{abstract}
\maketitle
%\end{CJK*} 

\section{Introduction}

Entropy plays a crucial role in various fields of physics, including but not limited to statistical mechanics, quantum information, many-body physics, and quantum chaos~\cite{Wehrl1978,Nielsen2012,Zeng2019,Huang1987,Han2015,Hu2019}. The Wehrl entropy was initially proposed by 
A. Wehrl in 1979 \cite{Wehrl1979}.
For 
a given quantum system with density operator ${\hat \rho}$, the Wehrl entropy is defined as the  Shannon entropy of the Husimi function
$\langle {\bm n}|{\hat \rho}|{\bm n}\rangle/Z$ \cite{Husimi1940}, where
$\{|{\bm n}\rangle\}$ are the coherent states of this system and $Z=\int d{\bm n}\langle {\bm n}|{\hat \rho}|{\bm n}\rangle$ serves as  the normalization constant.
For instance, for a single spin-1/2 particle,  $\{|{\bm n}\rangle\}$ are the SU(2) spin-coherent states, where  ${\bm n}$ is a  three-dimensional (3D) unit vector, {\it i.e.}, ${\bm n}\in S^2$ with $S^2$ being the unit sphere (Bloch sphere) of the 3D space.
Unlike the well-known von Neumann entropy, the Wehrl entropy is non-zero even for a pure state, and and can change with time during the evolution of a closed system. The Wehrl entropy has been applied and studied in various fields such as  quantum optics \cite{Orl/owski1995,Goes2020}, quantum information \cite{Mintert2004} and mathematical physics \cite{Lieb1978,Lee1988,Sugita2003,Lieb2014}. 
 Additionally, despite the coherent
states not forming a group of orthonormal bases, the Wehrl entropy can still be interpreted
as the entropy of a probability distribution of the outcome of a positive operator-valued measurement
(POVM) \cite{unpublished}.
%add a reference: C. Xu and P. Zheng, in preparation.

For the system of $N$  distinguishable spin-1/2 particles ($N>1$), there are  two different methods to define the Husimi function and the Wehrl entropy,  corresponding to different choices of the coherent states ${|{\bm n}\rangle}$. In the first method, 
${\bm n}\in S^2$
and $\{|{\bm n}\rangle\}$ are  SU$(2)$ coherent states, which are
multi-particle states with certain quantum numbers of 
the $N$-particle total spin and its component
along the direction ${\bm n}$
 \cite{Lee1988}. In the second method, $\{|{\bm n}\rangle\}$ are chosen 
to be the SU(2)$^{\otimes N}$ coherent states, which are the direct products of spin coherent states of each particle \cite{Sugita2003,Mintert2004}.
Consequently, in the second method, the coherent-state label ${\bm n}$ comprises $N$ components, with each one being a 3D unit vector, {\it i.e.}, ${\bm n}\in S^{2\otimes N}$.

In this work we investigate the Wehrl entropy of $N$ spin-1/2 particles, which are defined via the second method shown above. 
The Wehrl entropy, as defined by this method, has been previously explored for two spin-1/2 particles in \cite{Mintert2004}.
 Here we consider the systems with arbitrary particle number $N$, and focus 
on  the relationship between the Wehrl entropy and the  quantum entanglement.

 In 2003, A. Sugita \cite{Sugita2003} conjectured that  for this system the Wehrl entropy of all entangled states  are larger than that of the
  completely separable pure states. 
  He also proposed
%  Moreover, based on the definition of the corresponding Husimi function, Sugita \cite{Sugita2003} also 
that \cite{Sugita2003} 
  for many-body pure states
   the delocalization of the Husimi function, which can be measured by 
  the Wehrl entropy, reflect the complexity of quantum entanglement (entanglement complexity).
  However, to our knowledge, the Wehrl entropy 
  of   specific  entangled states of many-body systems (with particle number being  larger than 3)   has not been calculated in previous researches. 
  As a result, the connection between the Wehrl entropy and the entanglement complexity 
has not been examined and studied for specific examples of these many-body systems.

 Therefore,
in this work 
we  numerically calculate the Wehrl entropy for various pure states with respect to particle number $2\leq N\leq 20$. The Wehrl entropy  obtained by our calculation for the entangled pure states are all higher than that for the
the completely separable pure states, supporting  the above conjecture of A. Sugita \cite{Sugita2003}.
%Based on our results 
% we conjecture that the Wehrl entropy of all entangled pure states are larger than that of the
%  completely separable pure states,  {\it i.e.}, $\Lambda_LN$.  
% Therefore, we can define the entanglement degree of a multi-particle pure state as the the difference between its  WE  and $\Lambda_LN$. 
Furthermore, our numerical results indicate that
 in the large-$N$ ($N\gtrsim 10$) cases the Wehrl entropy per particle (WEPP) of  highly chaotic entangled states ({\it e.g},  $2^{-N/2}\sum_{s_1,s_2,...,s_N=\uparrow,\downarrow}|s_1,s_2,...,s_N\rangle e^{-i\phi_{s_1,s_2,...,s_N}}$, 
%or $\sum_{s_1,s_2,...,s_N=\uparrow,\downarrow}G_{s_1,s_2,...,s_N}|s_1,s_2,...,s_N\rangle$,  
with $\phi_{s_1,s_2,...,s_N}$ being independent random phases) are  significantly larger than those of  very regular entangled states, such as the Greenberger–Horne–Zeilinger (GHZ) state.

The above findings indicate that the Wehrl entropy effectively captures the entanglement complexity of the pure states of $N$ spin-1/2 particles. 
Due to this result and the fact that  the Wehrl entropy is invariant under local unitary transformations \cite{Sugita2003}, the WEPP serves as a quantitative measure of the  entanglement complexity of these states. Notably, unlike other metrics that assess entanglement complexity, such as those evaluating the entanglement between a subsystem and the rest of the particles, the WEPP does not require partitioning the entire system into two subsystems,   and thus may be suitable for many-body systems with long-range or all-to-all interactions, such as the Sachdev–Ye–Kitaev (SYK)  model with spin-1/2 particles \cite{sachdev1993}, quantum circuits \cite{hayden2007,sekino2008,shenker2015,lashkari2013,zhou2019}, and other various spin models \cite{georges2000,sherrington1975,derrida1980random,bray1980replica,thouless1977solution,parisi1979infinite,gardner1985spin,kirkpatrick1987dynamics,crisanti1992spherical,crisanti1993spherical,cugliandolo1993analytical,gross1985mean,kirkpatrick1987stable,sommers1981theory,usadel1987quantum,goldschmidt1990solvable,nieuwenhuizen1998quantum}.

%Our results yield that 
%the Wehrl entropy does reflect the entanglement complexity of many-body pure states. In particular,
% the Wehrl entropy per particle {\color{red}(WEPP)}  can be used as a 
% quantitative description of this entanglement complexity.
% Moreover, unlike  other quantities  used to evaluate this complexity, such as  the degree of entanglement between a subsystem and the other particles,  the Wehrl entropy per particle does not necessitate the division of the total system into two subsystems.
 
 Furthermore, our numerical results suggest that many-body pure states can be categorized into three types, each exhibiting distinct behaviors of WEPP as $N$ approaches infinity, and differing in entanglement complexity.
 Specifically, the states of one type exhibit highly chaotic entanglement, and the WEPP for $N\rightarrow\infty$   tends towards that of the maximum mixed state.
The states of another type are entangled very regularly, with the WEPP for $N\rightarrow\infty$  approaching the one of the completely separable pure states. The entanglement complexity and WEPP for $N \rightarrow \infty$ of the third type fall between those of the above two types.

% In addition to the above two problems, we also 
%numerically derive the time evolution of the Wehrl entropy of multi spin-1/2 particles governed by the time-dependent Schr\"odigner equation of various models, such as the Ising model with transverse and longitudinal fields.
%Our results 
%show that
%for the systems with a chaotic model and a Completely Separable initial state, the Wehrl entropy first  monotonically increase to a high-level region in a short time, and then oscillates in this region for long time.
%

The results of this work are helpful for 
the studies of the areas with entanglement 
complexity 
being very important, such as many-body dynamics of condensed-matter systems or quantum circuits, chaos and thermalization, spin liquids, and quantum computation, as well as the deep
understanding of  Husimi function and Wehrl entropy.

%The the statistical interpretation we developed for the  Wehrl entropy is not used
% in the discussions for the relationship between the Wehrl entropy and  quantum entanglement. Thus, the readers who are interested in the latter problem can skip the discussion for the former ({\it i.e.}, Sec.~\ref{pma}).

The remainder of this paper is organized as follows. In Sec.~\ref{def}
we introduce the definitions of the Husimi function and Wehrl entropy of $N$ spin-1/2 particles, and
% The  statistical interpretation of the Wehrl entropy is given in Sec.~\ref{pma}
  briefly outline some properties of the Wehrl entropy.
In Sec.~\ref{entangle}  
we investigate
the relationship between the Wehrl entropy and quantum entanglement.
Sec.~\ref{summary} contains a summary and some discussions. 
In the appendices we illustrate more properties of the Wehrl entropy, as well as
some details of our calculations.

\section{The Husimi Function and Wehrl Entropy of $N$  Spin-1/2 Particles}
\label{def}

\subsection{The Husimi Function}

We consider $N$   distinguishable spin-1/2 particles $1,...,N$.  An SU$(2)^{\otimes N}$ spin coherent state  is defined as a direct product of the spin coherent state of each particle:
\begin{eqnarray}
|{\bm {n}}\rangle\equiv|{{\bm n}}_1\rangle_1\otimes|{{\bm n}}_2\rangle_2\otimes...\otimes|{{\bm n}}_N\rangle_N,\label{sc}
\end{eqnarray}
with
\begin{eqnarray}
{\bm n}\equiv({{\bm n}}_1,{{\bm n}}_2,...,{{\bm n}}_N)\in S^{2\otimes N}.\label{nn}
\end{eqnarray}
 Here ${{\bm n}}_j \in S^2$ 
 ($j=1,...,N$) 
 is a 3D unit  vector, or the position vector of a point on the 
 unit sphere (Bloch sphere) $S^2$, 
 and can be expressed as ${\bm n}_j=(\sin\theta_j\cos\phi_j,\sin\theta_j\sin\phi_j,\cos\theta_j)$,
 with $\theta_j\in[0,\pi]$, $\phi_j\in[0,2\pi]$. Additionally, 
 $S^{2\otimes N}$ is the Cartesian product of $N$ unit spheres. Moreover, $|{{\bm n}}_j\rangle_j$ is the spin-coherent state of particle $j$ with respect to the direction ${{\bm n}}_j$, which satisfies 
\begin{eqnarray}
\left[{\hat {\bm \sigma}}^{(j)}\cdot  {{\bm n}}_j\right]|{{\bm n}}_j\rangle_j=|{{\bm n}}_j\rangle_j,
\end{eqnarray}
where ${\hat{\bm \sigma}}^{(j)}=(\hat{\sigma}_x^{(j)}, \hat{\sigma}_y^{(j)}, \hat{\sigma}_z^{(j)})$, with $\hat{\sigma}_{x,y,z}^{(j)}$ being the Pauli matrixes of the particle $j$  ($j=1,...,N$). The spin coherent states $\{|{\bm {n}}\rangle\}$ satisfy 
\begin{eqnarray}
\frac{1}{(2\pi)^N}\int d{\bm n}|{\bm n}\rangle\langle{{\bm n}}|= \hat I,\label{com}
\end{eqnarray}
with ${\hat I}$ being the identity operator, and
\begin{eqnarray}
\int d{\bm n}=\prod_{j=1}^N\int_0^\pi\sin\theta_jd\theta_j\int_0^{2\pi}d\phi_j.
\end{eqnarray}

The Husimi function $P_H({\hat \rho}; {\bm n})$ of this system is defined as
\begin{eqnarray}
P_H({\hat \rho};  {\bm n})\equiv\frac{1}{(2\pi)^N}\langle{{\bm n}}|{{\hat \rho}}|{\bm n}\rangle,\label{hf}
\end{eqnarray}
with ${{\hat \rho}}$ being the  density operator of the $N$-body quantum state, describing the quantum state of  all these particles. It is clear that $P_H({\hat \rho};  {\bm n})\geq 0$ for all ${\bm n}$, and
\begin{eqnarray}
\int d{\bm n}P_H({\hat \rho};  {\bm n})=1.
\end{eqnarray}
Notice that
 each  Husimi function $P_H({\hat \rho}; {\bm n})$ corresponds to a  unique $N$-body state ${\hat \rho}$. Namely,
 if ${\hat \rho}\neq{\hat \rho}^\prime$, then  there definitely exists ${\bm n}\in S^{2\otimes N}$, which satisfies $P_H({\hat \rho}; {\bm n})\neq P_H({\hat \rho}^\prime; {\bm n})$.

\subsection{The Wehrl Entropy}
\label{ppro}

Furthermore, the  Wehrl entropy  of these $N$ spin-1/2 particles 
 is defined as the  Shannon entropy corresponding to the  Husimi function:
\begin{eqnarray}
S_W({\hat \rho})\equiv-\int P_H({\hat \rho};  {\bm n})\ln\bigg[P_H({\hat \rho};  {\bm n})\bigg]d{\bm n}.\label{sw}
\end{eqnarray}
Clearly, the Wehrl entropy is a functional of the quantum state ${\hat \rho}$.

%It is clear that  this  WE 
%$S_W(|\psi\rangle\langle\psi|)$  is non-zero even  when the system is in a pure state.

%\section{Properties of the Wehrl Entropy}
%\label{ppro}

In Appendix~\ref{property}
we present and prove some properties of the Wehrl entropy and the Husimi function of $N$ spin-1/2 particles. Here we introduce two ones of them, which will be utilized in the subsequent sections.

{\bf Property 1:}  For all $N$-particle density operators ${\hat \rho}$, the Wehrl entropy per particle (WEPP) ({\it i.e.}, $S_W/N$) satisfies
\begin{eqnarray}
\frac {S_W({\hat \rho})}{N}\leq \Lambda_U,
\label{lambdaup}
\end{eqnarray}
with
\begin{eqnarray}
\Lambda_U\equiv\ln(4\pi)\approx 2.5310.
\label{lambdaup2}
\end{eqnarray}
The equality in Eq.~(\ref{lambdaup}) is satisfied when the system is in the state ${\hat \rho}= \bigotimes_{j=1}^N ({\hat I}_j/2)$, where ${\hat I}_j$ ($j=1,...,N$) is the identity operator for particle $j$. This state is commonly referred to as the ``maximum mixed state". This property is already known in previous researches.

{\bf Property 2:} The Wehrl entropy  $S_W({\hat \rho})$ is invariant under any local unitary transformation,  as proven in Ref. \cite{Sugita2003}. Explicitly,
 if ${\hat \rho}^\prime=U{\hat \rho} U^\dagger$ where
 $U=\bigotimes_{j=1}^N U_j$ and $U_j$ ($j=1,...,N$) is a unitary transformation acting on particle $j$, then  $S_W({\hat \rho})=S_W({\hat \rho}^\prime)$.
 This property implies that the Wehrl entropy is strongly related to quantum entanglement. 
 
% The above property leads to an important result on the Wehrl entropy of the completely
% separable states, {\it i.e.}, the  direct products of pure states of each particle.
% Since the particles are spin-1/2 ones, such a state is definitely a spin-coherent state $|{\bm n}\rangle$ defined in Eq.~(\ref{sc}).
% Every two such states  can be related via a local unitary transformation. Therefore,
% all the 
%  completely
% separable states 
% have the same WE, since 
%$S_W({\hat \rho})$ is invariant under any local unitary transformation.
%Explicitly, the direct calculations yield that 
% \begin{eqnarray}
% \frac{S_W(|{\bm n}\rangle\langle {\bm n}|)}{N}=\Lambda_L, \ \ \ \forall {\bm n}\in S^{2\otimes N},
%  \end{eqnarray}
%  where the constant $\Lambda_L$ being
%  \begin{eqnarray}
% \Lambda_L\equiv \frac 12+\ln(2\pi)\approx
% 2.3379.
%  \end{eqnarray}

% 
%  We first consider the cases where $|\psi\rangle$ is a $N$-body completely separable state, which is a direct product of the pure states of each particle.  Since the spin of each particle is 1/2, such a state is definitely a spin-coherent state $|{\bm n}\rangle$ defined in Eq.~(\ref{sc}).
%As shown in the above section, 
% $S_W({\hat \rho})$ is invariant under any local unitary transformation. Due to this fact, 

 \section{Wehrl entropy and  Entanglement}
 \label{entangle}

 In this section we focus on the Wehrl entropy of $N$-body pure states, 
 {\it i.e.}, $S_W(|\psi\rangle\langle\psi|)$, and investigate the relationship between the Wehrl entropy and quantum entanglement.

%  \begin{figure}[tbp]
%	\centering
%	\includegraphics[width=9cm]{pseudofig}
%	%\includegraphics[width=7cm]{pseudofig}
%        \caption{(color online.) \color{red}xxx}
%	\label{ent}
%\end{figure}
 
 %\subsection{The Wehrl entropy of Separable States}

% \subsection{Wehrl entropy of Completely Separable States}
% \label{wecss}
% 
%  We first consider the cases where $|\psi\rangle$ is a $N$-body completely separable state, which is a direct product of the pure states of each particle.  Since the spin of each particle is 1/2, such a state is definitely a spin-coherent state $|{\bm n}\rangle$ defined in Eq.~(\ref{sc}).
%As shown in the above section, 
% $S_W({\hat \rho})$ is invariant under any local unitary transformation. Due to this fact, all of the 
%  completely
% separable states have the same WE, since every two of them can be related via a local unitary transformation. 
%Explicitly, the direct calculations yield that 
% \begin{eqnarray}
% \frac{S_W(|{\bm n}\rangle\langle {\bm n}|)}{N}=\Lambda_L, \ \ \ \forall {\bm n}\in S^{2\otimes N},
%  \end{eqnarray}
%  where the constant $\Lambda_L$ is
%  \begin{eqnarray}
% \Lambda_L\equiv \frac 12+\ln(2\pi)\approx
% 2.3379.
%  \end{eqnarray}

   \subsection{Wehrl Entropy of Completely Separable Pure States}
 \label{wecss}
 
  We first consider the cases where $|\psi\rangle$ is a $N$-body completely separable pure state, {\it i.e.}, a direct product of pure states of each particle.  Since the spin of each particle is 1/2, such a state is definitely an $S^{2\otimes N}$ spin-coherent state $|{\bm n}\rangle$ defined in Eq.~(\ref{sc}).
  
As shown in the above section, 
 $S_W({\hat \rho})$ is invariant under any local unitary transformation. 
 On the other hand, every two completely
 separable states can be related via a local unitary transformation.
Therefore, all the 
  completely
 separable states have the same Wehrl entropy. 
Explicitly, the direct calculations yield that 
 \begin{eqnarray}
 \frac{S_W(|{\bm n}\rangle\langle {\bm n}|)}{N}=\Lambda_L, \ \ \ \forall {\bm n}\in S^{2\otimes N},
  \end{eqnarray}
 with
  \begin{eqnarray}
 \Lambda_L\equiv \frac 12+\ln(2\pi)\approx
 2.3379.
  \end{eqnarray}

  \subsection{Wehrl 
 Entropy of Two-Body Entangled States}
  \label{wetp}

  \begin{figure}[tbp]
	\centering
	\includegraphics[width=7.5cm]{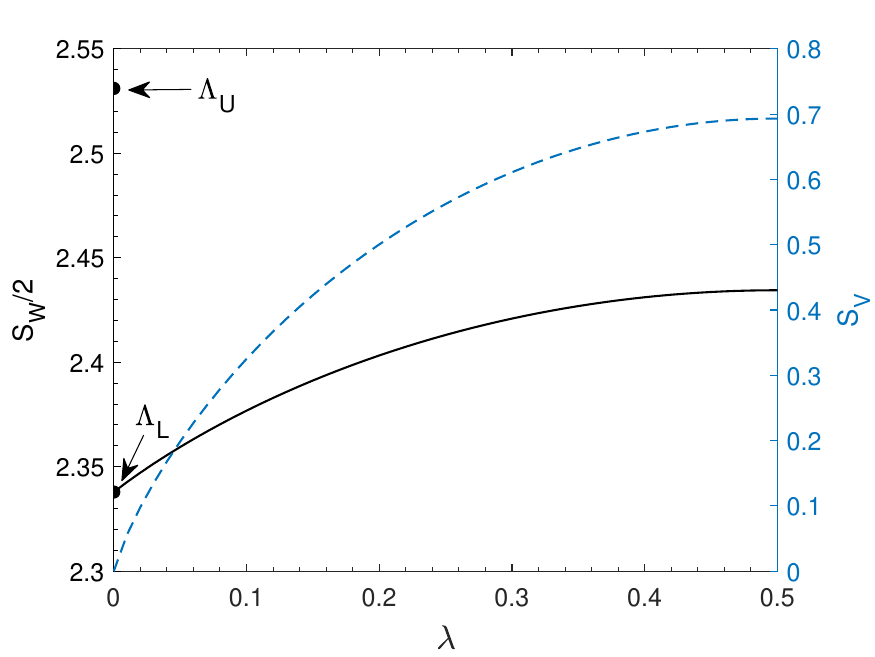}
	\caption{The WEPP of a two-body pure state $|\psi\rangle$  (black solid line) 
	and  the degree of entanglement $S_V(|\psi\rangle)$ (blue dashed line),
	as a functions of the parameter $\lambda$ defined in Sec.~\ref{wetp}. We also indicate the WEPP of the completely separable pure state and maximum mixed state, {\it i.e.}, $\Lambda_U$ and $\Lambda_L$. }
	\label{bound2b}
\end{figure}

 Using the Schmidt-decomposition technique, one can prove that each pure state $|\psi\rangle$ of two particles 1 and 2 can always be written as $|\psi\rangle=U_1\otimes U_2|\varphi\rangle$, where $U_j$ ($j=1,2$) is a unitary transformation acting on particle $j$, and 
 \begin{eqnarray}
 |\varphi\rangle=\sqrt{1-\lambda} |\uparrow\rangle_1 |\uparrow\rangle_2+\sqrt{\lambda}  |\downarrow\rangle_1 |\downarrow\rangle_2.\label{sd}
 \end{eqnarray}
Here $\lambda\in[0,1/2]$ is the smaller one of the two eigen-values of  the reduced density operator of particle 1, and $|\uparrow(\downarrow)\rangle_j$ ($j=1,2$) is the eigen-state of $\hat{\sigma}_z^{(j)}$ with eigen-value $+1(-1)$.
 Notice that  the degree of entanglement of  state $|\psi\rangle$ is characterized  by the von Neuman entropy of ${\hat \rho}_1$, {\it i.e.}, $S_V(|\psi\rangle\langle\psi|)\equiv-\lambda\ln\lambda-(1-\lambda)\ln(1-\lambda)$. For instance, $|\psi\rangle$ is a separable state
 when $\lambda=0$ ($S_V=0$), and is a maximum entanglement state (Bell state) when $\lambda=1/2$ ($S_V=\ln 2$).
 
%As demonstrated in Sec.~\ref{ppro}, the Wehrl entropy is invariable under local transformations. 
As mentioned in Sec.~\ref{ppro} (Property 2), the Wehrl entropy is invariant under local unitary transformations. Due to this fact,
we have $S_W(|\psi\rangle\langle\psi|)=S_W(|\varphi\rangle\langle\varphi|)$, and thus $S_W(|\psi\rangle\langle\psi|)$ is a function of the parameter $\lambda$.  In Fig.~\ref{bound2b} we illustrate 
 the WEPP
 $S_W(|\psi\rangle\langle\psi|)/N$ ($N=2$)  and the entanglement degree $S_V(|\psi\rangle\langle\psi|)$
 as functions of  $\lambda$.
 It is shown that  $S_W(|\psi\rangle\langle\psi|)/N$ increases with $\lambda$ or  $S_V(|\psi\rangle\langle\psi|)$. Therefore, the Wehrl entropy of a two-body pure entangled state is larger than the one of a separable state, and for all two-body pure states, the maximum entangled states ({\it i.e.}, the Bell states) have the largest Wehrl entropy.
 
%       \begin{figure}[tbp]
%	\centering
%	\includegraphics[width=8cm]{pseudofig}
%	\caption{(color online.) \color{red}xxx}
%	\label{2ent}
%\end{figure}

\subsection{Wehrl Entropy of Multi-Particle Entangled States}
\label{wemp}

Now we study the relationship between the Wehrl entropy and the entanglement of pure states of multi ($N\geq 3$) spin-1/2 particles.
Unfortunately,
even up to a local unitary transformation, one cannot express an arbitrary pure state of these particles  in a simple form like Eq.~(\ref{sd}), which has only one parameters. As a result, we cannot investigate the Wehrl entropy of every pure states, as done above for the two-particle systems. Alternatively,
we numerically calculate the Wehrl entropy for the entangled states of some typical types. 
%In this subsection we introduce the states we study and  the  results. Some further analysis  will be performed in the In the next subsection. 

We numerically calculate the Wehrl entropy for the following states (In the following the symbol $|s_1, s_2,\! ...,\! s_N\rangle$ ($s_1,s_2,...,s_N=\uparrow,\downarrow$) indicates the completely separable pure state $|s_1\rangle_1\otimes|s_2\rangle_2\otimes...\otimes|s_N\rangle_N$):
  %\begin{widetext}
\begin{itemize}
\item[$\bullet$] {\bf GHZ state}:
\begin{eqnarray}
|{\rm GHZ}\rangle\equiv\frac{1}{\sqrt{2}}\bigg[|\uparrow,\uparrow,...,\uparrow\rangle+|\downarrow,\downarrow,...,\downarrow\rangle     \bigg].\label{ghz}
\end{eqnarray}

\item[$\bullet$] {\bf W state}:
\begin{eqnarray}
|{\rm W}\rangle&\equiv&\frac{1}{\sqrt{N}}\bigg[|\downarrow,\uparrow,\uparrow,...,\uparrow\rangle
+|\uparrow,\downarrow,\uparrow,...,\uparrow\rangle\bigg.\nonumber\\
&&\ \ \ \ \ \ \bigg.+...+|\uparrow,\uparrow,...,\uparrow,\downarrow\rangle     \bigg].
\end{eqnarray}

        \begin{figure*}[tbp]
	\centering
			\includegraphics[width=10cm]{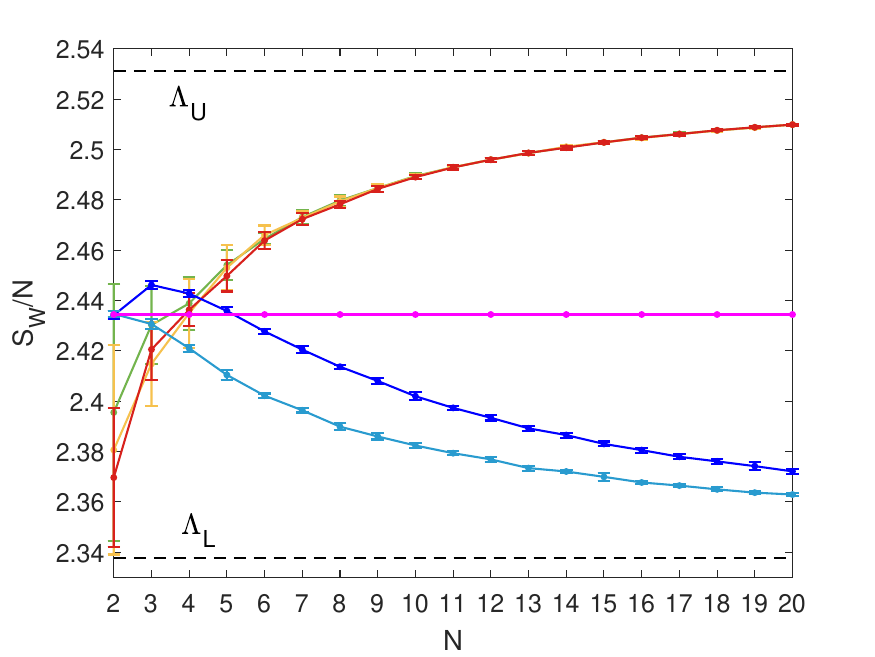}
			\includegraphics[width=1.7cm]{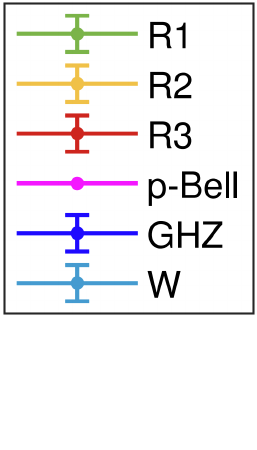}
	\caption{The WEPP of the states of types  defined in Sec.~\ref{wemp}. Here we show  $\langle S_W/N \rangle$ for the GHZ and W states, and $\langle\langle S_W/N \rangle\rangle$ for the R1, R2 and R3 states. The error bars indicates  $\delta_{\rm MC}$  of the the GHZ and W states and  $\delta_{\rm tot}$ of the R1, R2 and R3 states. We also show the analytical result of the WEPP of the p-Bell states without error bar. The WEPP of the completely separable pure state and maximum mixed state, {\it i.e.}, $\Lambda_U$ and $\Lambda_L$, respectively, are also shown as dashed lines. For more details, see Sec.~\ref{wemp}. 
	}
	\label{bound}
\end{figure*}

%with $H_{\rm Ising}$ being defined as
%\begin{eqnarray}
%H_{\rm Ising}&\equiv&-\sum_{j=1}^{N-1}\sigma_z^{(j)}\sigma_z^{(j+1)}-\sigma_x^{(N)}\sigma_x^{(1)}.
%\label{ising}
%\end{eqnarray}

\item[$\bullet$] {\bf p-Bell state}: the product-Bell state which is only defined only for even $N$, {\it i.e.},
\begin{eqnarray}
&&|{\rm p}{\text -}{\rm Bell}\rangle\equiv\nonumber\\
%\begin{array}{ll}
&&\bigotimes_{j=1,3,...,N-1}\left(|\uparrow\rangle_j|\uparrow\rangle_{j+1}+
|\downarrow\rangle_j|\downarrow\rangle_{j+1}
\right)/\sqrt{2}.
\label{h}
\end{eqnarray}

\item[$\bullet$] {\bf R1-state}: 
the state with equal amplitude and random sign, {\it i.e.},
\begin{eqnarray}
&&|{\rm R1}\rangle\equiv \nonumber\\
&&\frac{1}{\sqrt{2^N}}\sum_{s_1,s_2,...,s_N=\uparrow,\downarrow}
(-1)^{\xi_{s_1,s_2,...,s_N}}
|s_1,s_2,...,s_N\rangle,\nonumber\\
\label{R2}
\end{eqnarray}
where  $\xi_{s_1,s_2,...,s_N}$ $(s_j=\uparrow,\downarrow;j=1,...,N)$ are $2^N$ independent random coefficients, with each one taking values $0$ or $1$ with equal probabilities.

\item[$\bullet$] {\bf R2-state}: the state with
``equal amplitude and random phase", {\it i.e.},
\begin{eqnarray}
&&|{\rm R}2\rangle\equiv \nonumber\\
&&\frac{1}{\sqrt{2^N}}\sum_{s_1,s_2,...,s_N=\uparrow,\downarrow}|s_1,s_2,...,s_N\rangle e^{-i\phi_{s_1,s_2,...,s_N}},\nonumber\\
\label{R2}
\end{eqnarray}
where  $\phi_{s_1,s_2,...,s_N}$ $(s_j=\uparrow,\downarrow;j=1,...,N)$ are $2^N$ independent random angles, with each one taking values in the region $[0,2\pi)$ with constant probabilistic density $1/(2\pi)$.

\item[$\bullet$] {\bf R3-state}: the ``totally-random" state, {\it i.e.},
\begin{eqnarray}
&&|{\rm R}3\rangle\equiv \nonumber\\
&&\frac 1Z
\!\!
\sum_{s_1, s_2, ..., s_N=\uparrow,\downarrow}\!\!\!\!\!\!
C_{s_1, s_2, ..., s_N}
|s_1, s_2, ..., s_N\rangle e^{-i\phi_{s_1, s_2, ..., s_N}},\nonumber\\
\label{R2}
\end{eqnarray}
where $C_{s_1, s_2,\! ...,\! s_N}$ and $\phi_{s_1,s_2,...,s_N}$ $(s_j=\uparrow,\downarrow;j=1,...,N)$ are $2^N$ independent random positive numbers and independent  random angles, respectively.
Explicitly, each $C_{s_1, s_2,\! ...,\! s_N}$ taking values in the region $[0,1]$ with constant probabilistic density $1$, and each $\phi_{s_1,s_2,...,s_N}$ taking values in the region $[0,2\pi)$ with constant probabilistic density $1/(2\pi)$, and $Z=\sqrt{\sum_{s_1, s_2, ...,s_N=\uparrow,\downarrow}
C_{s_1, s_2, ..., s_N}^2
}.$

\end{itemize}

We analytically calculate the Wehrl entropy  for the p-Bell states,
and numerically calculate those
for other states
by  performing the integration in Eq.~(\ref{sw}) using the Monte Carlo method, for $2\leq N\leq 20$. Explicitly,
for the GHZ and W states, we perform the  integration for ten times  for each particle number $N$, and then derive the average value $\langle S_W/N \rangle$  as well as the standard deviation $\delta_{\rm MC}$. For the R1, R2 and R3 states, for each $N$ 
we first generate five samples of the  random parameters in the definitions of these states ({\it i.e.}, the $\xi$-, $C$- and $\phi$-parameters). For each state with respective to a certain sample, we
calculate $\langle S_W/N \rangle$ and  $\delta_{\rm MC}$ as above. We further derive the average value 
and the standard deviation   of $\langle S_W/N \rangle$ 
of these five states, which are denoted as $\langle\langle S_W/N\rangle\rangle$  and $\delta_{\rm R}$, respectively. Additionally, we define  $\delta_{\rm tot}\equiv\delta_{\rm R}+\delta_{\rm MC}^{\rm max}$, where $\delta_{\rm MC}^{\rm max}$ is the 
 maximum  value of $\delta_{\rm MC}$ 
 these five states. 
 
 In Fig.~\ref{bound} we show the WEPP
 obtained from the above calculations as functions of $N$.
 Explicitly, we illustrate $\{ \langle S_W/N \rangle,\delta_{\rm MC}\}$ for the
 GHZ and W states, and  illustrate $\{\langle\langle S_W/N\rangle\rangle, \delta_{\rm tot}\}$ for the R1, R2 and R3 states. The analytical results of $S_W/N$ for the p-Bell states is also shown in this figure.

In addition to the aforementioned states, we also calculate the  Wehrl entropy of the time-dependent states $|\psi(t)\rangle$ governed by the time-dependent Schr\"odinger equation of some models listed in Appendix~\ref{dynamical models}.

From the these calculations, we obtain the following understandings for the Wehrl entropy of multi-particle entangled pure states:

\subsubsection*{\bf (a) The lower and upper bounds of Wehrl entropy}
\label{suba}

As shown in Fig.~\ref{bound}, the WEPP of all the states of types (i-v) are larger than the one of the completely separable pure states, {\it i.e.}, $\Lambda_L$. The same is true for the WEPP of we obtained for the states of  Appendix~\ref{dynamical models}.   These results are consistent with the conjecture of 
	A. Sugita  \cite{Sugita2003}, which yields that for our system the Wehrl entropy takes the minimum value for the spin coherent states.

According to this conjecture, $\Lambda_L$ is the lower bound of the WEPP of 
all density operators \cite{conjecture2002}.

%To our knowledge, the conjecture that  $\frac {S_W(\hat \rho)}{N}\geq \Lambda_L$ has not been analytically proven for arbitrary state $\hat \rho$. 
%In Ref.~\cite{Lieb2014} it is demonstrated that this conjecture is correct for any symmetric pure state. Additionally, 
%In 2002 A. Sugita \cite{Sugita2002} proved that this the result corresponding to this conjecture 
%for
%$
%S_W^{(q)}({\hat \rho})\equiv\ln\big[\int P_H({\hat \rho};  {\bm n})^qd{\bm n}\big]/(1-q)
%$. $S_W^{(q)}({\hat \rho})$ with $q=2,3,4,...$.}

This result, together with  the property shown in Eq.~(\ref{lambdaup}), further leads to 
\begin{eqnarray}
\Lambda_L\leq \frac {S_W({\hat \rho})}{N}\leq \Lambda_U,\ \ \ \ {\rm for}\ \ \forall {\hat \rho}.
\label{bounds}
\end{eqnarray}

\subsubsection*{\bf (b) Wehrl entropy and entanglement complexity   of multi-particle pure states}

Now we investigate the relation between the Wehrl entropy and entanglement complexity, with the help of the results shown in Fig.~\ref{bound}. 

%For our system, each $N$-body pure state $|\psi\rangle$ can be expanded in the basis $\{|s_1,s_2,...,s_N\rangle| s_1,s_2,...,s_N=\uparrow,\downarrow\}$ as $|\psi\rangle=\sum_{s_1,s_2,...,s_N=\uparrow,\downarrow}
%c_{s_1,s_2,...,s_N}
%|s_1,s_2,...,s_N\rangle$, and the entanglement complexity of $|\psi\rangle$
%refers to the degree of disorder of the coefficients $\{c_{s_1,s_2,...,s_N}\}$.
%According to this definition, the entanglement complexity of would be much higher than the 
%one of the GHZ/W states, because he R1/R2/R3 states include $2^N$ terms with very random coefficients, while the GHZ or W states only include 2 or $N$ terms with equal coefficients. 
%
%On the other hand, as shown in Fig.~\ref{bound}, when the particle number $N$ is large enough ($N\gtrsim 10$)
%the {\color{red}WEPP} of the 
%the R1, R2 or R3 states, which are almost same, are significantly  larger than the ones of the GHZ or W states.  
%These results yield that
%the Wehrl entropy  and  the  entanglement 
% complexity are  positively correlated. That is consistent with the following facts: 
% According to the aforementioned meaning of entanglement 
% complexity,
% when the entanglement
% complexity of a state is larger, the  Husimi function $P_H({\bm n})$ of this state would be more 
% disordered. On the other hand,
% the Wehrl entropy is mathematically defined as the Shannon entropy of the Husimi function $P_H({\bm n})$, which just describes how disorder the Husimi function. Thus, the Wehrl entropy increases with the entanglement 
% complexity.
%}

As shown in Fig.~\ref{bound}, when the particle number $N$ is large enough ($N\gtrsim 10$.)
the Wehrl entropy per particle of the 
the R1, R2 or R3 states, which are almost same, are significantly  larger than the ones of the GHZ or W states.  
On the other hand, 
we notice that there is big difference between the R1/R2/R3 states and the GHZ/W states:  the former ones 
are very chaotic (complicated) while the latter ones are very regular (simple). 
Explicitly,  the R1/R2/R3 states include $2^N$ terms with very random coefficients, while the GHZ or W states only include 2 or $N$ terms with equal coefficients. 
Thus, our result yield that
the Wehrl entropy  is larger when  $|\psi\rangle$ is entangled more complicated. 
This is consistent with the following two facts:  (I) When the entanglement
  of a state is more complicated, the  Husimi function $P_H({\bm n})$ of this state would be more 
 disordered. (II)
 The Wehrl entropy is mathematically defined as the Shannon entropy of the Husimi function $P_H({\bm n})$, which just describes how disordered the Husimi function is. Thus, the Wehrl entropy increases with the entanglement 
 complexity.

Moreover, since the Wehrl entropy is invariant under local unitary transformations (the property 2 of Sec.~\ref{ppro}), for pure states it is only determined by the entanglement.
Therefore,  one can use the WEPP as a quantitative description of 
the entanglement complexity of $N$-body pure states of spin-1/2 particles.

%This is a big difference between the Wehrl entropy and some
%  other descriptions of  the  entanglement  complexity of   a multi-particle system, {\it e.g.}, the entanglement entropy between a subsystem and the other particles. 
   In the researches for many-body problems of spin particles, people have already used
  some  descriptions of the complexity of entanglement, ({\it e.g.}, the entanglement entropy between a subsystem and the other particles).
 These descriptions require dividing the total system into two subsystems. They are suitable for the systems with short-range inter-particle interactions, as one can naturally consider particles in close proximity as a subsystem. 
   
 However, there are also some important models or systems with all-to-all interactions, such as the Sachdev–Ye–Kitaev (SYK)  model with spin-1/2 particles \cite{sachdev1993}, quantum circuits \cite{hayden2007,sekino2008,shenker2015,lashkari2013,zhou2019}, and other various spin models \cite{georges2000,sherrington1975,derrida1980random,bray1980replica,thouless1977solution,parisi1979infinite,gardner1985spin,kirkpatrick1987dynamics,crisanti1992spherical,crisanti1993spherical,cugliandolo1993analytical,gross1985mean,kirkpatrick1987stable,sommers1981theory,usadel1987quantum,goldschmidt1990solvable,nieuwenhuizen1998quantum}. For these models and systems, there is no inherent reason to classify specific particles into one subsystem while placing others into another subsystem. Thus, these descriptions of entanglement complexity are not suitable.
On the other hand, in the calculation of Wehrl entropy, one does not need to divide the system into two subsystems. Therefore, the WEPP provides an effective description of entanglement complexity for systems with all-to-all interactions. The WEPP can be used to, {\it e.g.}, quantify the growth of entanglement during quantum dynamics and thus providing insights into how the system approaches quantum thermalization \cite{srednicki1994chaos,deutsch1991quantum,abanin2017rigorous,kuwahara2016floquet}, or detection of potential entanglement transitions in systems subjected to repeated measurements \cite{li2018quantum,cao2019entanglement,li2019measurement,skinner2019measurement,chan2019unitary,bao2020theory,choi2020quantum,gullans2020dynamical,gullans2020scalable,tang2020measurement,buchhold2021effective,sang2021measurement,lavasani2021measurement}.

%[one can use WEPP to ... application...(low entropy...scar...time evolution...)]}

%As mentioned above, we calculate the  Wehrl entropy of the time-dependent states $|\psi(t)\rangle$  of the models  in Appendix~\ref{dynamical models}. Our results show that when $N\geq 10$ the {\color{red}WEPP} of all these states are below or equal to the ones of the R1/R2/R3 states.
   %\end{widetext}

%Since the entanglement degree ${\mathbb E}$ can describe the ``chaoticity" of the quantum entanglement  of a mulit-particle pure state, and can be calculated efficiently for a given state $|\psi\rangle$, it would be helpful for the studies related to the ``chaoticity" of a pure multi-particle entangled state, {\it e.g.}, the quantum thermalization. 

\subsubsection*{\bf (c) Behaviors of Wehrl entropy in the limit $N \rightarrow\infty$}

  \begin{figure}[tbp]
	\centering
	\includegraphics[width=7cm]{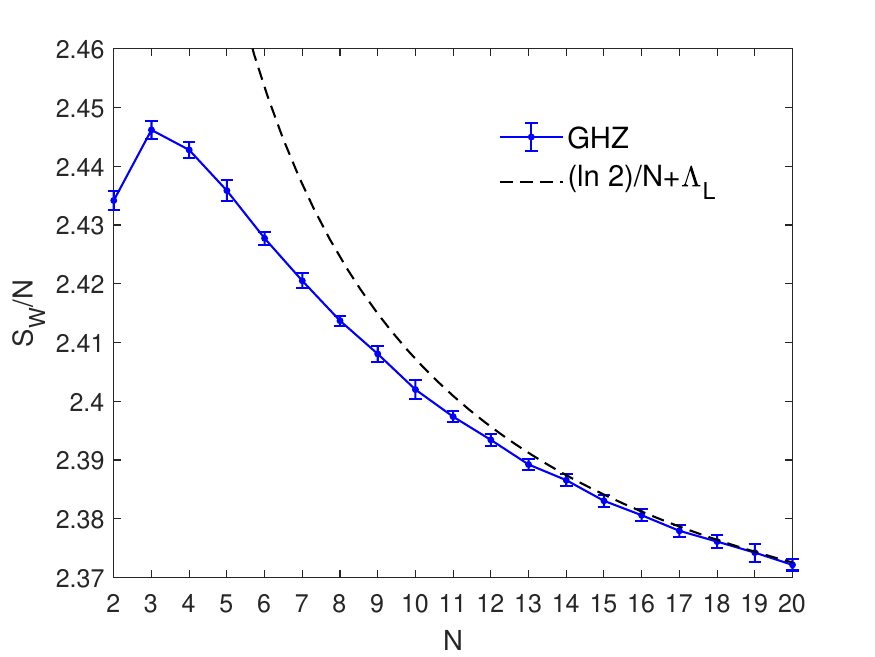}
        \caption{{\bf Solid line with error bars}: the WEPP of the GHZ states, which are given by our numerical calculation ({\it i.e.}, the results shown in Fig.~\ref{bound}). {\bf Dashed line}:  $(\ln 2)/N+\Lambda_L$. }
	\label{ghz}
\end{figure}

Now we consider the behaviors of the WEPP in the large-$N$ limit.

Fig.~\ref{bound} shows that when the particle number $N$ is large,
 the WEPP of the R1, R2 and R3  states increase with  $N$, while the one 
 of the ones of GHZ/W states decrease with $N$.  We numerically 
 fit the results of these states
  with $N\geq16$ as functions of $N$, and find that within the error bar of the fitting, in the limit $N\rightarrow\infty$ the WEPP of the R1/R2/R3  states 
 and the ones of  the GHZ/W states
 tend $\Lambda_U$ and $\Lambda_L$, respectively. The details of the fitting results are shown in the footnote \cite{fitting}. 
Due to these results, we conjecture that
 \begin{eqnarray}
\lim_{N\rightarrow\infty} \frac{S_W}N&=&\Lambda_U\ \ \ {\rm   for\  R1/R2/R3\ states},\label{r1}\\
\lim_{N\rightarrow\infty} \frac{S_W}N&=&\Lambda_L \ \ \ {\rm for \ GHZ/W\ states}.\label{r2}
 \end{eqnarray}
Moreover,
 in Appendix~\ref{limit} 
 %we analytically prove that $\lim_{N\rightarrow\infty} {S_W}/N=\Lambda_L$ for the GHZ state, and 
  we performs  rough analysis which supports the above conjectures for  the GHZ and R1/R2 states.
  Additionally, this analysis also show that for GHZ state we have $S_W({\hat \rho})/N\approx (\ln 2)/N+\Lambda_L$ for the large-$N$ cases. As shown in Fig.~\ref{ghz}, this approximate expression agrees well with our numerical results when $N\gtrsim 16$. Additionally, it can be directly proved that  the WEPP of the 
 p-Bell states is a  $N$-dependent constant (approximately 2.43) between $\Lambda_L$ and $\Lambda_N$, as shown in Fig.~\ref{bound}. 
 
Here we emphasis that both the conclusions (\ref{r1}) and (\ref{r2}) are not intuitive. These two results indicate that, in the large-$N$ limit, the upper and lower bounds of the WEPP for {\it pure states} are identical to those for {\it all density operators}, a surprising outcome given the typically distinct nature of pure and mixed states.  Furthermore, while the GHZ state is widely recognized as a maximally entangled state, its entanglement complexity is unexpectedly low, so that the WEPP of the GHZ state approaches that of separable states as $N \rightarrow \infty$.

     \begin{figure}[tbp]
	\centering
	\includegraphics[width=8.5cm]{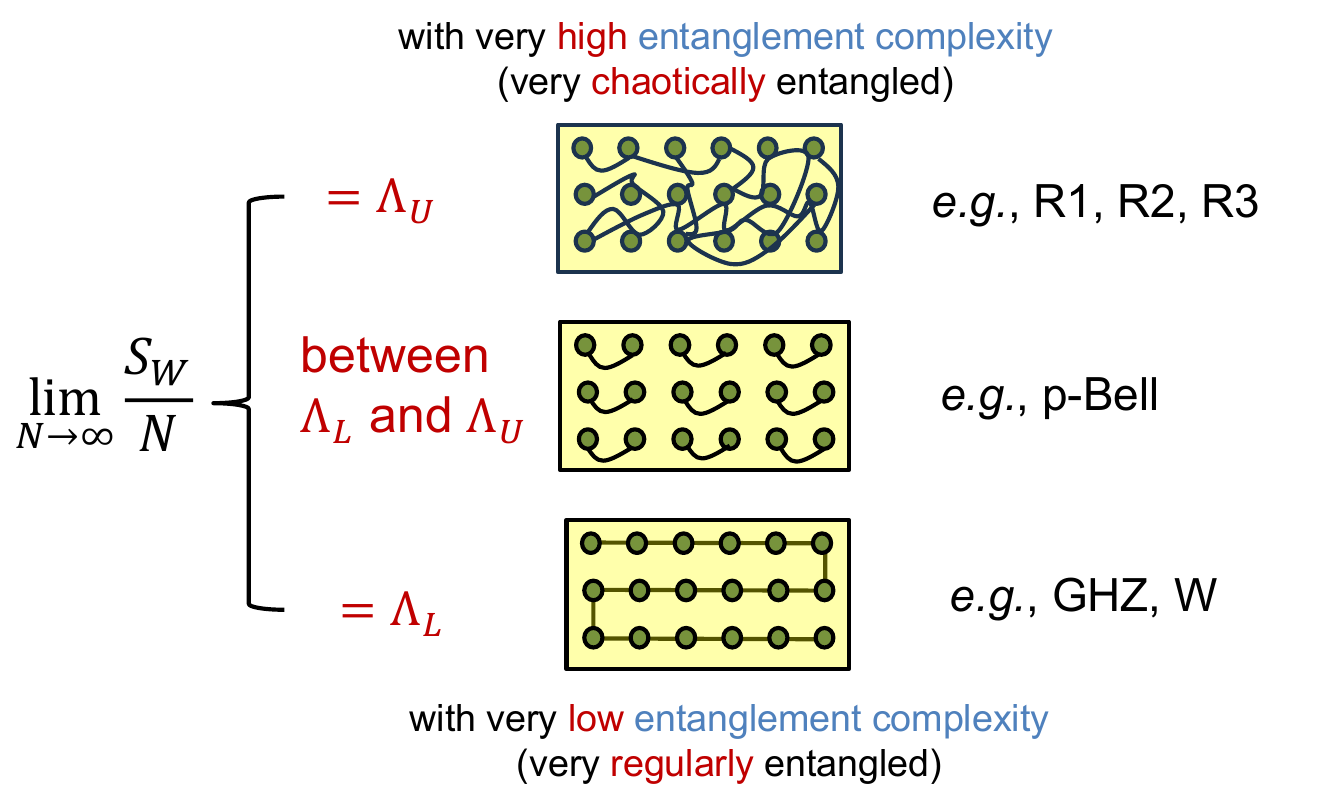}
        \caption{
        Schematic diagrams of the  three types of many-body pure states with different behaviors of $\lim_{N\rightarrow\infty}S_W/N$.
        The detailed discussions are given in Sec.~\ref{wemp}.}
	\label{3type}
\end{figure}

 Our above results yield that the multi-particle entangled pure states of our system can be classified into three types, according to the behaviors of $S_W/N$ (Fig.~\ref{3type}), {\it i.e.}, the states with:
 \begin{itemize}
 \item[] $ \lim_{N\rightarrow\infty}\frac {S_W} N=\Lambda_U$\ \ \  \ \ \ \ \ \ \ \ \ \  (type I),
  \item[] $\Lambda_L<\lim_{N\rightarrow\infty}\frac {S_W} N<\Lambda_U$\ \ \ \ \  (type II),
   \item[] $\lim_{N\rightarrow\infty}\frac {S_W} N=\Lambda_L$\ \ \  \ \ \ \ \ \ \ \ \ \  (type III).
 \end{itemize}
 Specifically, states of type I ({\it e.g.}, the R1/R2/R3 states) exhibit highly chaotic entanglement, resulting in such a high entanglement complexity that, for $N\rightarrow\infty$ the Wehrl entropy tends towards that of the maximum mixed state ({\it i.e.}, the upper bound of the Wehrl entropy for all density operators).
 In contrast, the states of type III ({\it e.g.}, the GHZ/W states) are entangled very regularly, so that for $N\rightarrow\infty$ the Wehrl entropy tends to the one of the completely separable pure states  ({\it i.e.}, the lower bound of the Wehrl entropy for all density operators). The entanglement complexity of the states of the type II are between that of the above two types.
In Fig.~\ref{3type}, we present schematic diagrams of these three types of quantum states.

\section{Summary and Discussions}
\label{summary}

In this work we evaluate the relation between the Wehrl entropy and  entanglement of quantum pure states
of $N$ spin-1/2 particles. 
Our results yield that  the WEPP  can be used as a quantitative description for the entanglement complexity of multi-particle pure states, without dividing the total system into two or more subsystems. 
Thus, WEEP is particularly 
 suitable 
 for systems with  all-to-all interactions. 
%In this work we calculate the Wehrl entropy via directly performing the integration in Eq.~(\ref{sw}) analytically or numerically. 
%It is possible to develop other efficient numerical approaches for the calculation of WE, {\it e.g.},
%expanding the function $\ln \big[P_H({\hat \rho};  {\bm n})\big]$ in Eq.~(\ref{sw}) as series of $1-P_H({\hat \rho};  {\bm n})$ and then doing the integration individually for each term.
%The strong relation between the Wehrl entropy and quantum entanglement is essentially due to the fact that the $N$-body coherent states $\{|{\bm n}\rangle\}$ defined in Eq.~(\ref{sc}) actually includes all completely-separable states. Therefore, 
%in the generalization of our results to the systems with $N$ spin-$J$ particles ($J>1/2$), the coherent states are still defined as in  Eq.~(\ref{sc}), and the $1$-body coherent state $|{\bm n}_j\rangle_j$ of the particle $j$ ($j=1,...,N$) takes on all 
%normalized pure states of this spin-$J$ particle (up to a global phase factor), including but not limited to SU$(J)$ spin coherent states. 
We further show that many-body pure states can be categorized into three types, {\it i.e.}, the types I, II and III shown above, according to the behavior of $\lim_{N\rightarrow\infty}S_W/N$. The entanglement complexity differs among states of various types.  

The aforementioned results are beneficial for the investigation
of quantum many-body systems in complicated many-body entangled states, including condensed-matter
systems and quantum circuits, and thus are helpful for
the studies for the related problems, such as thermalization
and spin-liquids.

To our knowledge, the Wehrl entropy cannot be directly measured in current experiments. In the future, it would be both interesting and important to establish connections between the Wehrl entropy power potential (WEPP) and observable physical effects in many-body systems, such as investigating the distinct, measurable properties of the three types of quantum states mentioned above. Furthermore, the classification based on the behavior of $\lim_{N\rightarrow\infty} S_W/N$ is well supported by the examples presented in this work. Developing more rigorous mathematical frameworks in this context would be highly beneficial. 
Additionally, extending the results of this work to indistinguishable particles is a worthwhile avenue for further exploration.

%\end{itemize}

\begin{acknowledgments}
The authors thank Yingfei Gu, Dazhi Xu, Pengfei Zhang, Hui Zhai, Honggang Luo, Zhiyuan Xie, and Ninghua Tong  for fruitful discussions.
This work was supported by the National Key Research and Development Program of China (Grant  No. 2022YFA1405300), and 
NSAF Grant No. U1930201.
\end{acknowledgments}

\bigskip

\appendix

\section{Properties of the Wehrl Entropy and Husimi Function}
\label{property}

In this appendix, we demonstrate and prove some properties of the Husimi function and the Wehrl entropy of $N$ spin-1/2 particles, including but not limited to the two ones presented in Sec.~\ref{ppro}.

\subsection{The Relation Between the Husimi Functions of the Total System and  Subsystems }

If the  density operator of $N$ spin-1/2 particles is  ${\hat \rho}$, then the reduced density operator ${\hat \rho}_{\rm sub}$ of
 a subsystem including $M$ particles $i_1,i_2,..,i_M$ is
\begin{eqnarray}
{\hat \rho}_{\rm sub} ={\rm Tr}_{i\notin\{i_1,i_2,..,i_M\}}{\hat \rho}.
\end{eqnarray}
Here ${\rm Tr}_{i\notin\{i_1,i_2,..,i_M\}}$ means tracing over the particles expect $i_1,i_2,..,i_M$.
The Husimi function $P_H({\hat \rho}_{\rm sub};  {\bm n}_{\rm sub})$, with ${\bm n}_{\rm sub}\equiv({{\bm n}}_{i_1},{{\bm n}}_{i_2},...,{{\bm n}}_{i_M})\in S^{2\otimes M}$, is related to the Husimi function $P_H({\hat \rho};  {\bm n})$ of all the $N$-particles via 
\begin{eqnarray}
P_H({\hat \rho}_{\rm sub};  {\bm n}_{\rm sub})=\int P_H({\hat \rho};  {\bm n})\prod_{i\notin\{i_1,i_2,..,i_M\}}  d{\bm n}_i .
\label{phsub}
\end{eqnarray}
This can be proven directly with the definition of the Husimi function.

\subsection{The Subadditivity, Monotonicity and Concavity of the Wehrl Entropy}

The $N$ spin-1/2 particles can be separated into two subsystems $A$ and $B$, with $A$ including particles $i_1,i_2,..,i_M$ and $B$ including other particles. If the density operators of the $N$ particles is ${\hat \rho}$, and the reduced operators of the subsystems $A$ and $B$ are ${\hat \rho}_A$ and ${\hat \rho}_B$, respectively, then we have
\begin{eqnarray}
 S_W({\hat \rho})\leq  S_W({\hat \rho}_A)+S_W({\hat \rho}_B)=S_W({\hat \rho}_A\otimes{\hat \rho}_B)\nonumber\\
  \ \  \ \  \ \  \ \  \ \  \ \  \ \ \ \ {\rm (subadditivity),}\label{subadd}
\end{eqnarray}
and
\begin{eqnarray}
 S_W({\hat \rho}_A),\ S_W({\hat \rho}_B)\leq S_W({\hat \rho}) \ \ \ \ {\rm (monotonicity).}
\end{eqnarray}
Furthermore, if three $N$-body density operators  ${\hat \rho}$, ${\hat \rho}^{(1)}$ and ${\hat \rho}^{(2)}$ satisfy ${\hat \rho}=p_1{\hat \rho}^{(1)}+p_2{\hat \rho}^{(2)}$, with $p_{1,2}> 0$ and $p_1+p_2=1$, when we have
\begin{eqnarray}
 S_W({\hat \rho})\geq p_1S_W({\hat \rho}^{(1)})+p_2 S_W({\hat \rho}^{(2)}) \ \ \ \ {\rm (concavity).}\nonumber\\
\end{eqnarray}
These results can be proven directly via Eq.~(\ref{phsub}) and the methods used in Ref.~\cite{Wehrl1978} (subadditivity and concavity) and Ref.~\cite{Wehrl1979} (monotonicity).

Due to Eq.~(\ref{subadd}), for every $N$-body state ${\hat \rho}$, we have 
\begin{eqnarray}
 S_W({\hat \rho})\leq  \sum_{j=1}^{N}S_W({\hat \rho}_j),\label{swrapp}
\end{eqnarray}
with ${\hat \rho}_j$ ($j=1,...,N$) being the reduced density operator of particle $j$. On the other hand, it can be directly proved that for a spin-1/2 particle, the Wehrl entropy is at most $\Lambda_U\equiv \ln(4\pi)$. Using this fact and Eq.~(\ref{swrapp}), we obtain the ``property 1" of Sec.~\ref{ppro}.

\section{Dynamical Models}
\label{dynamical models}

As mentioned in Sec.~\ref{wemp},
we have calculated the  Wehrl entropy of the states $|\psi(t)\rangle$ determined by  the
 time-dependent Schr\"odinger equation with the following Hamiltonians.

\begin{itemize}
\item[(1):] The Ising model with Hamiltonian
\begin{eqnarray}
	H_{\rm Ising}&\equiv&-\left[ \sum_{j=1}^{N-1}\sigma_z^{(j)}\sigma_z^{(j+1)}+\Bigg.\sigma_z^{(N)}\sigma_z^{(1)}\right].
	\label{ising}
\end{eqnarray}
For this model we consider the cases with initial state 
\begin{eqnarray}
|\psi(t=0)\rangle=\bigotimes_{j=1}^N(|\uparrow\rangle_j+|\downarrow\rangle_j)/\sqrt{2},\label{xstate}
\end{eqnarray}
particle number $N=5$, and evolution time $t\in[0,40]$, as well as that with the same initial state and 
$2\leq N\leq 14$, $t\in[30,40]$.
 
\item[(2):] The XY model with Hamiltonian
\begin{eqnarray}
H_{\rm XY}&\equiv&-\Bigg\{\sum_{j=1}^{N-1}\left[\sigma_x^{(j)}\sigma_x^{(j+1)}+\sigma_y^{(j)}\sigma_y^{(j+1)}\right]+\Bigg.\nonumber\\ 
&&\ \ \ \ \Bigg.\sigma_x^{(N)}\sigma_x^{(1)}+\sigma_z^{(N)}\sigma_z^{(1)}\Bigg \}.\label{hxy}
\end{eqnarray}
The initial states, particle numbers, and evolution times of the cases we consider for this model are same as those of the above Ising model.

%For this model we consider the cases with the initial state of Eq.~(\ref{xstate}), 
%
%
%
%$N=2,3,\cdots,14;\quad t=30-40$\\
%$N=15;\quad t=0-40$

\item[(3):] The Ising model with both a transverse and a longitudinal field with Hamiltonian
\begin{eqnarray}
H_{\rm TLI}\equiv H_{\rm Ising}-\sum_{j=1}^N\left[\sigma_y^{(j)}+\sigma_z^{(j)}\right],
\label{tli}
\end{eqnarray}
with $H_{\rm Ising}$ being defined in Eq.~(\ref{ising}).
The initial states, particle numbers, and evolution times of the cases we consider for this model are same as those of the above Ising model.

% $|\psi(t=0)\rangle=\bigotimes_{j=1}^N|{\bm e}_x\rangle_j=\bigotimes_{j=1}^N(|\uparrow\rangle_j+|\downarrow\rangle_j)/\sqrt{2}$
%
%$N=2,3,\cdots,14;\quad t=30-40$\\
%$N=15;\quad t=0-40$

\item[(4):] The Ising model with both a transverse and a longitudinal field with Hamiltonian
\begin{eqnarray}
	H_{\rm TLI}\equiv H_{\rm Ising}-\sum_{j=1}^N\left[\sigma_x^{(j)}+g_z\sigma_z^{(j)}\right],
	\label{tli}
\end{eqnarray}
with $H_{\rm Ising}$ being defined in Eq.~(\ref{ising}) and $g_z=0,0.1,0.5,1,2$. For this model we consider 
the cases with particle number $N=6$, evolution time $t\in[0,40]$ and
the initial states being either the one of Eq.~(\ref{xstate}) or $|\psi(t=0)\rangle=\bigotimes_{j=1}^N|{\bm u}_j\rangle_j$, where
${\bm u}_j(j=1,\cdots,N)$ are random directions.
\end{itemize}

\section{Behaviors of $S_W/N$ for $N\rightarrow\infty$}
\label{limit}

In this appendix we provide two rough analysis, which imply that   for the R1 and  R2 states
we have $\lim_{N\rightarrow\infty}S_W/N=\Lambda_U$,
and for the GHZ states we have $\lim_{N\rightarrow\infty}S_W/N=\Lambda_L$ and $S_W/N\approx (\ln 2)/N+\Lambda_L$ in the large-$N$ limit.

We first consider the GHZ states defined in Eq.~(\ref{ghz}). According to Eq.~(\ref{hf}), the Husimi function of this state is given by
\begin{eqnarray}
P_H({\bm n})=\frac{1}{2}\left\{P_H^{(1)}({\bm n})+P_H^{(2)}({\bm n})+P_H^{(3)}({\bm n})+P_H^{(3)}({\bm n})^\ast\right\}, \nonumber\\
\label{phghz}
\end{eqnarray}
where $P_H^{(1)}({\bm n})$ and $P_H^{(2)}({\bm n})$ are the Husimi functions of the states $|\uparrow,\uparrow,...,\uparrow\rangle$ and $|\downarrow,\downarrow,...,\downarrow\rangle$, respectively, and
\begin{eqnarray}
P_H^{(3)}({\bm n})=\frac{1}{(2\pi)^N}\langle {\bm n}|\uparrow,\uparrow,...,\uparrow\rangle\langle \downarrow,\downarrow,...,\downarrow|{\bm n}\rangle.
\label{p3}
\end{eqnarray}
Furthermore, the straightforward calculations yield that 
\begin{eqnarray}
P_H^{(1)}({\bm n})
%=\prod_{j=1}^N\frac{\vert \langle{\bm n}_j|\uparrow\rangle\vert^2}{2\pi}
&=&\prod_{j=1}^N
\frac{\cos(\theta_j/2)^2}{2\pi};\label {ph1}\\
P_H^{(2)}({\bm n})
%=\prod_{j=1}^N\frac{\vert \langle{\bm n}_j|\uparrow\rangle\vert^2}{2\pi}
&=&\prod_{j=1}^N
\frac{\sin(\theta_j/2)^2}{2\pi};\label {ph2}\\
\vert P_H^{(3)}({\bm n})\vert
%=\prod_{j=1}^N\frac{\vert \langle{\bm n}_j|\uparrow\rangle\vert^2}{2\pi}
&=&\prod_{j=1}^N
\frac{\sin(\theta_j/2)\cos(\theta_j/2)}{2\pi},\label {ph3}
\end{eqnarray}
where  $\theta_j\in[0,\pi]$ is the polar angle of ${\bm n}_j$, with ${\bm n}\equiv({{\bm n}}_1,{{\bm n}}_2,...,{{\bm n}}_N)$, as shown in Sec.~\ref{def}. Eqs~(\ref{ph1}-\ref{ph3}) show that when $N\rightarrow\infty$,
the functions $P_H^{(1)}({\bm n})$ and $P_H^{(2)}({\bm n})$ significantly  at the points with $\theta_1=\theta_2=...=\theta_N=0$ and  $\theta_1=\theta_2=...=\theta_N=\pi$, respectively. So we suppose that the contributions of $P_H^{(1,2)} ({\bm n})$ to the integral of the expression (\ref{sw}) of the Wehrl entropy are mainly given by that of the regions around these two peaking points. 
Moreover, the maximum value of $\vert P_H^{(3)}({\bm n})\vert
$  ($(4\pi)^{-N}$) is much less  the peaking values of $P_H^{(1,2)} ({\bm n})$ ($(2\pi)^{-N}$). Therefore, in the calculation of the Wehrl entropy with Eq.~(\ref{sw}) we totally ignore the contributions from $P_H^{(3)}({\bm n})$.
The similar analysis yields that in this calculation one can also 
 also ignore $P_H^{(1)}({\bm n})$ in the regions around the peaking point of $P_H^{(2)}({\bm n})$, and vice versa. Thus, in the limit $N\rightarrow\infty$ the Wehrl entropy of the GHZ states is 
\begin{eqnarray}
S_W({\hat \rho})&\approx&-\frac 12\int P_H^{(1)}({\hat \rho};  {\bm n})\ln\bigg[P_H^{(1)}({\hat \rho};  {\bm n})/2\bigg]d{\bm n}\nonumber\\
&&-\frac 12\int P_H^{(2)}({\hat \rho};  {\bm n})\ln\bigg[P_H^{(2)}({\hat \rho};  {\bm n})/2\bigg]d{\bm n}\nonumber\\
&=&\ln 2+\Lambda_LN, \label{appghz}
\end{eqnarray}
which yields 
$\lim_{N\rightarrow\infty}S_W({\hat \rho})/N=\Lambda_L$ and $S_W/N\approx (\ln 2)/N+\Lambda_L$ in the large-$N$ limit.

Now we consider the R2 states defined in Sec.~\ref{wemp}. The density operator ${\hat \rho}=|{\rm R2\rangle\langle \rm R2|}$
of this state can be expressed as a density matrix in the basis $\{|s_1, s_2, ..., s_N\rangle, s_1,s_2,...,s_N=\uparrow,\downarrow\}$ the density operator ${\hat \rho}=|{\rm R2\rangle\langle \rm R2|}$. The diagonal elements 
of this density matrix
are all $1/2^N$. The norm of the non-diagonal elements are also $1/2^N$, but the complex phases of the non-diagonal elements are very random when $N\rightarrow\infty$, due to the random phases $\phi_{s_1,s_2,...,s_N}$ in the expression~(\ref{R2}) of this state. On the other hand, in the calculations of the Wehrl entropy with Eq.~(\ref{sw}), there are summations for the these non-diagonal matrix elements. Due to these random phases we suppose that these summations can be ignored in the limit $N\rightarrow\infty$, and thus in the calculation of the Wehrl entropy one can only keep the diagonal elements of the density matrix, {\it i.e.}, make the approximation $S_W({\hat \rho})\approx S_W({\hat \rho}^\prime)$, with ${\hat \rho}^\prime=
\sum_{s_1,s_2,...,s_N=\uparrow,\downarrow}|s_1, s_2, ..., s_N\rangle\langle s_1, s_2, ..., s_N|/(2\pi)^{N/2}
$. It is clear that  ${\hat \rho}^\prime$ is just the density operator of the maximum mixed state, {\it i.e.}, ${\hat \rho}= \bigotimes_{j=1}^N ({\hat I}_j/2)$, where ${\hat I}_j$ ($j=1,...,N$) is the identity operator for particle $j$. Thus, we have $S_W({\hat \rho})/N \approx \Lambda_U$. 

Moreover, an analysis similar to above  also implies that $S_W({\hat \rho})/N \approx \Lambda_U$ for the R1 states.

%\begin{widetext}
\bibliography{reference.bib} % code #2

\begin{thebibliography}{59}
\bibitem [{con()}]{conjecture2002}%
  \BibitemOpen
  \href@noop {} {}\bibinfo {note} {To our knowledge, the conjecture that
  $S_W(\hat \rho)/N\geq \Lambda_L$ has not been analytically proven for
  arbitrary state $\hat \rho$. In Ref.~\cite{Lieb2014} it is demonstrated that
  this conjecture is correct for any symmetric pure state. Additionally, In
  2002 A. Sugita \cite{Sugita2002} proved that this the result corresponding to
  this conjecture for $ S_W^{(q)}({\hat \rho})\equiv\ln\big[\int P_H({\hat
  \rho}; {\bm n})^qd{\bm n}\big]/(1-q) $. $S_W^{(q)}({\hat \rho})$ with
  $q=2,3,4,...$.}\BibitemShut {Stop}%
\bibitem [{fit()}]{fitting}%
  \BibitemOpen
  \href@noop {} {\ }\bibinfo {note} {Explicitly, we use the fitting function
  $A+B/N+C/N^2$. Thus, the value of $A$ given by the fitting is the Wehrl
  entropy per particle in the limit $N\rightarrow\infty$.The fitting results we
  obtained are $A=2.53313\pm 0.00444$ for the R1/R2/R3 states, $A=2.33226\pm
  0.01798$ for the GHZ states, and $2.34337 \pm 0.01782$ for the W
  states.}\BibitemShut {Stop}%
\end{thebibliography}

\end{document}